\documentclass[11pt]{article}
\pdfoutput=1
 \usepackage{mciteplus}
 \usepackage{tikz}
 \usepackage{color}
 \usepackage{xcolor}
 \usepackage{comment}
 \definecolor{darkblue}{rgb}{0.1,0.1,.7}
 \usepackage[colorlinks, linkcolor=darkblue, citecolor=darkblue, urlcolor=darkblue, linktocpage,hyperfootnotes=false]{hyperref} 
\usepackage{epsfig}
\usepackage{graphicx}
\usepackage{cite}
\usepackage{amsfonts}
\usepackage{amssymb}
\usepackage{bm}
\usepackage{latexsym}
\usepackage{mathtools}
\usepackage{amsmath}
\numberwithin{equation}{section}
\def\bq{\begin{quote}}
\def\eq{\end{quote}}
 at 10truept

\newcommand{\cala}{{\cal A}}
\newcommand{\calc}{{\cal C}}

\newcommand{\call}{{\cal L}}

\newcommand{\beq}{\begin{equation}}
\newcommand{\eeq}{\end{equation}}
\newcommand{\beqa}{\begin{eqnarray}}
\newcommand{\eeqa}{\end{eqnarray}}
\newcommand{\bea}{\begin{eqnarray}}
\newcommand{\eea}{\end{eqnarray}}

\def\roughly#1{\raise.3ex\hbox{$#1$\kern-.75em\lower1ex\hbox{$\sim$}}}

\begin{document}

\thispagestyle{empty}
\begin{titlepage}
  \bigskip

  \bigskip\bigskip

  \bigskip

\begin{center}
{\Large \bf {Quantum gravity observables: observation, algebras, and  mathematical structure
}}
    \bigskip
\bigskip
\end{center}

  \begin{center}

 \rm {Steven B. Giddings\footnote{\texttt{giddings@ucsb.edu}} }
  \bigskip \rm
\bigskip

{Department of Physics, University of California, Santa Barbara, CA 93106, USA}  \\
\rm

  \bigskip \rm
\bigskip
 
\rm

\bigskip
\bigskip

  \end{center}

\vspace{3cm}
  \begin{abstract}

The questions of describing observables and observation in quantum gravity appear to be centrally important to its physics.  A relational approach holds significant promise, and a classification of different types of relational observables (gravitationally dressed, field relational, and more general) is outlined.  Plausibly 
gravitationally dressed observables are particularly closely tied to the fundamental structure of the theory.  These may be constructed in the quantum theory to leading order in Newton's constant, and raise important questions about localization of information.  Approximate localization is given by a ``standard dressing" construction of a ``gravitational splitting."  It is also argued that such gravitational dressings give a generalization of the crossed product construction, reducing to this and yielding type II von Neumann algebras in special cases.  Gravity therefore introduces a significantly more general alteration of the algebraic structure of local quantum field theory, also with apparent connections to holography, but whose  implications have not been fully understood.  In particular, properties of the algebra of gravitationally dressed observables suggest a possible role for other non-algebraic structure on the Hilbert space for quantum gravity.

 \medskip
  \noindent
  \end{abstract}
\bigskip \bigskip \bigskip 

\noindent{\it Dedicated to two leaders in quantum gravity: Jim Hartle, who also collaborated on and helped inspire work towards understanding its observables, and Stanley Deser, whose work laid important foundations for development of the subject.}

  \end{titlepage}

\section{Introduction}

The subjects of observations and observables in quantum gravity appear to play a profoundly important role in formulating the theory.  They are multifaceted, and touch on the basic questions ranging from description of observation from within a quantum system to that of the fundamental mathematical structure of the theory.

In fact, the word ``observable" has different meanings to different people, sometimes leading to confusion.  A first use of the word is to describe what observers or experimentalists actually measure.  Examples here include gravitational wave signals as measured in interferometers, or temperature fluctuations in the microwave sky. Another use of the word is to describe what can in principle be measured, for example the S-matrix or vector, or inclusive cross sections.  Finally, the word observable is also used to describe certain operators in a quantum theory, with properties such as hermiticity and gauge invariance.  The latter play a key role in formulation of quantum theories such as quantum field theory, and quite possibly also do so in quantum gravity.  To help disambiguate the different meanings, we sometimes use the term\cite{GiSl} ``q-observable" in this last context, when helpful.

At the level of physics, clearly these different meanings are related, although the precise relation is an important question for further investigation. The purpose of this piece is primarily to describe some aspects of q-observables, although this of course touches on questions associated with the other two meanings.

To summarize the structure of the paper, the next section will review the important role of subalgebras of observables in describing the localization of information, and a notion of subsystems, in quantum field theory.  Section three then turns to the problem of local gauge invariant observables in gravitational theories, describing an approach to localization via relational observables.  A classification of different types of relational observables -- gravitationally dressed, field relational, and more general -- is described, with discussion of some examples.  Section four discusses construction of gravitationally dressed observables, first in the classical context, and then extending to quantum operators, specifically in a weak-gravity expansion.  Section five describes how such gauge invariant operators significantly modify the algebraic structure from that of field theory, and describe an approach to approximate localization of information via a gravitational splitting, obtained by constructing ``standard dressings."  Section six argues that standard dressing constructions generalize constructions of a crossed product algebra, which have played a role in exhibiting type II von Neumann algebras in gravity.  Section seven describes  further how the ultimate algebraic structure arising from gravity apparently involves a much more significant deviation from that of field theory than the type II algebras, and argues that the fundamental mathematical structure of quantum gravity may in fact go beyond that of algebras.

\section{Observables and algebras in quantum field theory}
\label{LQFTsec}

Before considering gravity, it is instructive to review aspects of the key role of observables in local quantum field theory (LQFT).  Observables (here, meaning q-observables) in LQFT are centrally important for describing the property of {\it locality} - which may be described both as {\it localization of information}, and as {\it local propagation of information}.  The basic idea is that we have some local observables, say including $O(x)$ and $O'(x)$, in the theory.  We can then use the former to construct a state whose excitation  is localized to a neighborhood $U$ of a fixed background spacetime,
\beq\label{Jstate}
|\psi_J\rangle=e^{-i\int dx J(x){ O}(x)}|0\rangle
\eeq
where $J(x)$ is a source whose support lies in $U$, and $|0\rangle$ is a fiducial vacuum state.  Then locality is described by stating that
\beq
\label{statloc}
\langle\psi_J|{ O}'(x')|\psi_J\rangle = \langle0|{O}'(x')|0\rangle
\eeq
for a point $x'$ spacelike separated from $U$, meaning that the observable at $x'$ is insensitive to the excitation of the non-trivial state.   If we work perturbatively in $J$, this arises from $[O(x),O'(x')]=0$ for $x'$ spacelike to $x$. If $x$ and $x'$ lie in a common spacelike surface, {\it e.g} that with time $t=0$ in Minkowski space, this describes localization of information to the region.  In relativistic quantum field theory, the Poincar\'e symmetry then relates this to the more general statement of commutativity for unequal times, which is the restriction of local propagation (no faster than light signaling) in the theory.

In fact, this notion of localization of information gives a general way to describe {\it subsystems} in LQFT.  If we have two spacelike separated neighborhoods, $U$ and $U'$, then the subalgebras of observables localized to the respective regions, $\cala_U$ and $\cala_{U'}$, will commute.  Such subalgebras then give us a way to construct independent quantum states, by generalizing \eqref{Jstate}.  These subalgebras play a foundational role in a careful formulation of the theory.\footnote{See, {\it e.g.}, \cite{Haag} for a more complete description of this, in the framework of algebraic quantum field theory.}  Since they are associated with neighborhoods of the spacetime, they also have inclusion and overlap relations; they form a ``net" that mirrors the net of open sets, which in turn describes the topological structure of the manifold.  In this sense, the structure of the background spacetime manifold is encoded in the quantum mechanical theory precisely through the algebra of observables and this net of subalgebras.  

This discussion may seem formal, but appears profoundly {\it physically} important.  This is illustrated by a quote from no less than A.~Einstein:\footnote{See \cite{Einsep}, and for a translation \cite{Einstr}.}
``... it appears to be essential for this arrangement of the things introduced in physics that, at a specific time, these things claim an existence independent of one another, insofar as these things `lie in different parts of space.' Without such an assumption of the mutually independent existence (the `being-thus') of spatially distant things, an assumption which originates in everyday thought, physical thought in the sense familiar to us would not be possible. Nor does one see how physical laws could be formulated and tested without such a clean separation."  This passage describes a notion of what has been called Einstein separability.  In the preceding discussion, this ``mutually independent existence" is encoded in the commutativity of the subalgebras.

\section{Gravity and relational observables}

Once we include gravity, things appear to be fundamentally different, in a way we have so far not fully understood in a quantum formulation of the theory.

A starting point for the problem is the fact that with gravity, there are {\it no} local observables.  This was begun to be understood earlier, but was formulated in a general way by Torre in \cite{Torr}.  The basic problem is that a local scalar observable $O(x)$ will transform under an infinitesimal diffeomorphism parameterized by $\xi^\mu$ as 
\beq
\delta_\xi{ O}(x)=-\xi^\mu \partial_\mu{ O}(x)\ .
\eeq
But, in gravity the diffeomorphisms are gauge transformations, so $O(x)$ is not gauge invariant; the problem also extends to more general observables that are localized in any particular region.  This is a starting point for the question of how to describe observables and observation in gravitational theory.  This is apparently one of the fundamental questions in quantum gravity.

A very general proposal for constructing (q-)observables is to construct {\it relational} observables.  These represent a large class of observables, and have a long history.\footnote{For a limited set of references in this long history, see \cite{Koma,DewiG,Rove1991,Marotime,Ditt,GMH,Tamb,SGalg,DoGi1,Frob:2017gyj,GHK}.}  The basic idea is to solve the preceding gauge invariance problem by {\it localizing with respect to other features of the state.}

Beginning with classical fields, suppose in $D$ dimensions we have $D$ functionals $X^A[\phi_I,g_{\mu\nu}](x)$
of the quantum fields $\phi_I$ and metric $g_{\mu\nu}$, also depending on coordinates $x$, and transforming as scalars under diffeomorphisms acting on the fields, {\it i.e.}
\beq
\delta_\xi X^A(x) = -\xi^\mu\partial_\mu X^A(x)\ .
\eeq
Suppose also that we are in a configuration of the fields and metric where the condition $X^A(x) =y^A$ specifies a point, 
\beq\label{invX}
x^\mu=\chi^\mu(y^A)\ , 
\eeq
given by the inverse function, which varies with parameters $y^A$.  Then
\beq\label{brobs}
{ O}(\chi^\mu(y^A))
\eeq 
is {\it diffeomorphism invariant} and gives a basic example of a relational observable; it can be referred to as $O$ at the point $x$ where $X^A=y^A$.  Clearly this construction can be extended to more general localized observables, {\it e.g.} multi-local observables.

There are different such constructions to localize relationally, and it's helpful to introduce some terminology to distinguish them.  For example, in the case where our ``localizers" only depend on the fields but not the metric, $X^A[\phi_I](x)$, a natural nomenclature is that the observable   \eqref{brobs}  is ``field relational" (or, if preferred, ``matter relational").  The case where the localizers only depend on the metric, $X^A[g_{\mu\nu}](x)$ is captured by the terminology ``gravitationally dressed."  These both are then special cases of the general, mixed, observable constructed using $X^A[\phi_I,g_{\mu\nu}](x)$.\footnote{Some other nomenclature has more recently emerged in the literature, see {\it e.g.} \cite{CLPW}.  There, observables defined with respect to an idealized observer, which might be thought of as an idealization of our field relational construction, are referred to as ``gravitationally dressed."  However, while it is true that these are constructed so as to respect  some of the diffeomorphisms (see further discussion in Sec.~\ref{crosssec}), which do play the role of gauge transformations of gravity, that does not necessarily mean they are associated with any dynamical gravitational field, unlike what under the present definitions are termed gravitationally dressed or mixed observables.  The former could equally well be defined in a theory with {\it no} dynamical metric. Further discussion of relations between observables appears in \cite{GHK}.}

A very simple example of purely field relational observables is provided by the $Z$-model of \cite{GMH}, with antecedents in \cite{DewiG}.  Suppose that $Z^A(x)$ are $D$ scalar fields, and consider a state\footnote{If we generalize to a quantum state, this could be true as a statement about expectation values.} where they have values $Z^A(x)=\lambda \delta^A_\mu x^\mu$.  Then the observable $O(x)$ evaluated at the point where $Z^A(x)=y^A$ can be constructed as
\beq
 \int d^D x\Big\vert\frac{\partial Z^A}{\partial x^\mu} \Big\vert\ \prod_A\delta(Z^A(x)-y^A)\ O(x)\ .
\eeq

\begin{figure}[h]
 	\begin{center}
 		\includegraphics[width=0.40\textwidth]{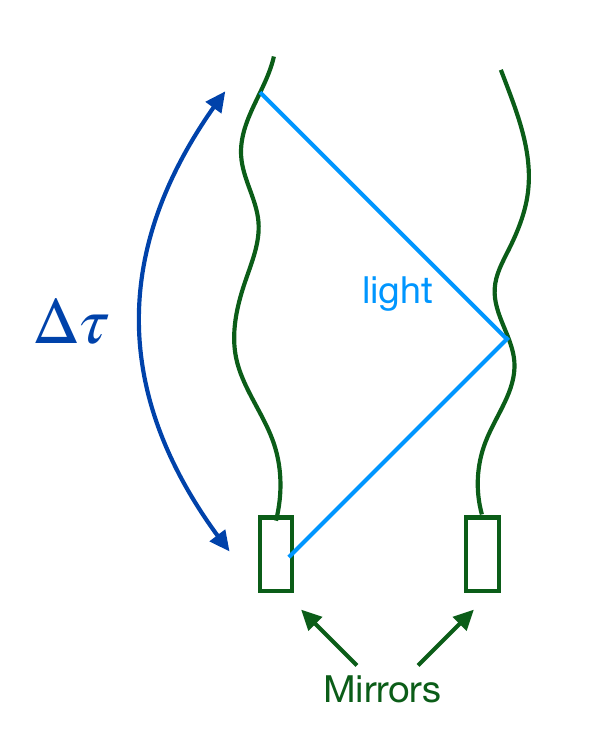} 
 		\caption{Schematic of an interferometer, in which measurements of the proper time of light propagation depends on the gravitational perturbations in the vicinity of the interferometer.}
 		\label{fig-interfer}
 	\end{center}
 \end{figure}

More generally, we might  expect that field relational observables, or mixed observables, are in some cases directly connected to observation, since we might think of them as for example localizing $O(x)$ with respect to a detector or observer.  For example, one kind of observation in gravity, using gravitational interferometers, is idealized by considering the motion of two mirrors, and measuring the time $\Delta \tau$  a light pulse emitted from one mirror takes to bounce off the other and return, as in Fig.~\ref{fig-interfer}.  One can then consider properties of, for example, $\langle\psi|\Delta\tau\, \Delta \tau |\psi\rangle$ in the corresponding state.  Constructing the corresponding state and quantum observables is still a little complicated, and part of the general open question of how to formulate observables and relate them to observations in such observing situations, although some recent progress and discussion of the interferometer case appears in \cite{LeZu,CKS}.

This also raises another general point, whose full impact has yet to be fully understood, that we expect observers to be quantum subsystems (possibly in an approximation) that are part of the bigger system.  This for example seems to place limits on observational precision, due to their quantum nature (for simple examples, see \cite{GMH,GaGiobs}).  A contrasting viewpoint has been postulated, in which certain ``ideal observers" are part of the fundamental description of the physics\cite{Wittobs}, but it remains to be seen whether such  ideal observers exist as part of a full mathematical description of the theory.

The quantum version of this general discussion of relational observables also makes contact with extensive literature on related questions under a different name, that of quantum reference frames; see \cite{GHK,DEHK}, also for more extensive references.  If we can define quantum operators such as \eqref{brobs} (as we will begin to see, there are nontrivial questions in such a definition), then the reference variables or locators $\chi^\mu(y^A)$ can be thought of as associated to such a quantum reference frame.

\section{Gravitationally dressed observables}\label{GDO}

In contrast to more complicated situations involving matter observers, gravitationally dressed observables provide a ``minimal" construction of relational observables, which begin with an observable of QFT and give minimal additional gravitational structure necessary for gauge invariance.  This also suggests that they connect to questions regarding the fundamental structure of quantum gravity, and in particular the question of how information is localized in a gravitational system, where the non-trivial dressing has apparently important implications.

 \begin{figure}[h]
 	\begin{center}
 		\includegraphics[width=0.55\textwidth]{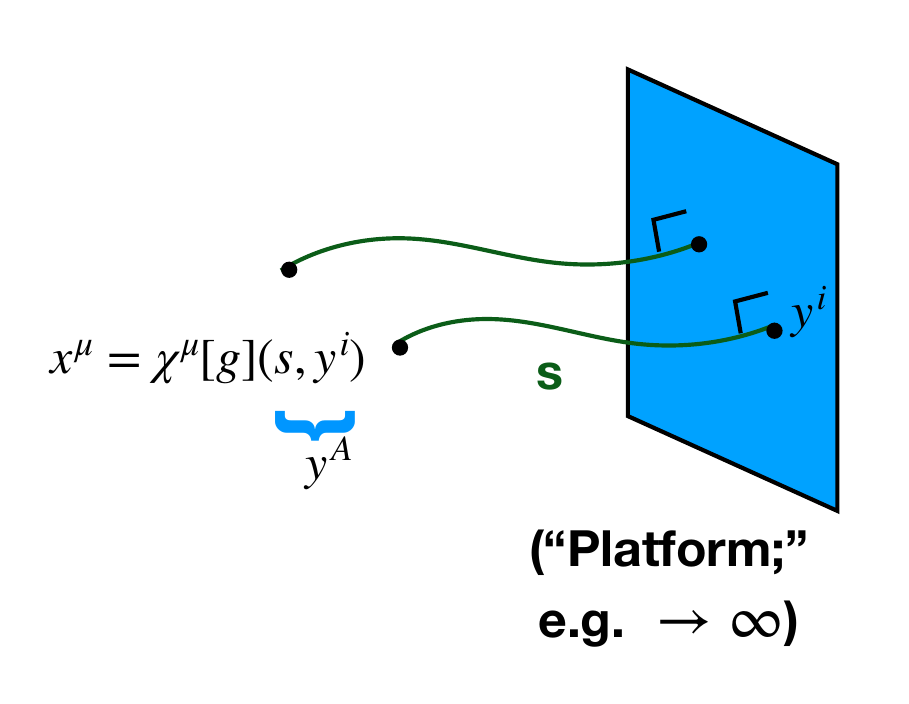} 
 		\caption{Illustration of construction of a gravitational dressing based on a platform.  A point $x$ may be located using geodesics launched perpindicular to the platform, and specified in terms of the platform coordinate $y^i$ and geodesic distance $s$.}
 		\label{fig-platform}
 	\end{center}
 \end{figure}

A simple version of gravitationally dressed observables can be given beginning in the classical case, following \cite{DoGi1}.\footnote{Earlier related constructions include \cite{Heem} and \cite{KaLigrav}. The first of these derived related nontrivial commutators
as arising from the constraints, but didn't give the dressed operators, and the second focussed on
deriving commuting bulk operators. Further discussion of such observables also appears in \cite{GHK}.}
Consider a ``platform," {\it e.g} at a fixed large $x^1$, as illustrated in Fig.~\ref{fig-platform}.  We can consider taking it to infinity, since diffeomorphisms there are typically {\it not} considered to be gauge; or, an example in the case of anti de Sitter space is to take the platform to lie at the AdS boundary\cite{GiKi}.  
Then, a general point can be specified by finding the geodesic launched perpindicularly to the platform that goes through the point; the coordinate of the geodesic origin on the platform, $y^i$, and the proper distance, $s$, determine the point.  Analogously to \eqref{invX},
\beq
x^\mu=\chi^\mu[g](s,y^i)\ ,
\eeq
where here $s, y^i$ play the role of $y^A$.  The construction clearly depends on the metric, and is diffeomorphism invariant under diffeomorphisms not acting at the platform.
For example if we consider a theory with a minimally-coupled scalar $\phi(x)$, which is a basic QFT observable, the corresponding gravitationally dressed version of it is
\beq
\Phi(y^A) = \phi(\chi^\mu[g](s,y^i))\ ;
\eeq
we refer to this construction as providing a gravitational line dressing.
Such constructions can be extended to dress fields with higher spin, {\it e.g.} by using the $\chi^\mu[g](s,y^i)$ to construct frames, and so we have a broad class of gravitationally dressed observables in classical field theory.

A key question, then, is how to extend such constructions to the quantum theory.   The complete answer remains an open problem, but we appear able to make progress by treating gravity perturbatively,\footnote{An alternate approach gives more complete constructions in the simple case of lower-dimensional gravity\cite{HaWuobs}, although these may not capture important features of higher-dimensional theories.}
using the expansion ({\it e.g.} with flat background metric $\eta_{\mu\nu}$)
\beq
g_{\mu\nu}=\eta_{\mu\nu}+\kappa h_{\mu\nu}\ ,
\eeq
with $\kappa=\sqrt{32\pi G}$.  For example, in such a perturbative expansion, we can find the preceding geodesic functions $\chi^\mu$ to leading order\cite{DoGi1,QGQF}. They are
\beq
\chi^\mu(y)=y^\mu+ V^\mu[h](y)
\eeq
where
\beq\label{linedress}
V_\mu(y)= {\kappa\over 2} \int_y^\infty dx^{\nu} \left\{ h_{\mu\nu}(x) + \int_{x}^\infty dx^{\prime\lambda}\left[\partial_\mu h_{\nu\lambda}(x') - \partial_\nu h_{\mu\lambda}(x')\right]\right\} + {\cal O}(\kappa^2)\ ,
\eeq
with integrals along the direction $x^1$ transverse to the platform.
Then, also to leading perturbative order in $\kappa$, 
\beq\label{Phisum}
\Phi(y)= \phi(y) + V^\mu(y)\partial_\mu\phi +{\cal O}(\kappa^2)\ .
\eeq
This is easily seen to be diffeomorphism invariant to leading order.  This follows from the transformation law
\beq\label{keyc}
\delta_{\kappa\xi}\phi =-\call_{\kappa \xi}\phi = -\kappa\xi^\mu\partial_\mu\phi\ ,
\eeq
together with the ``key relation" 
\beq\label{Kreln}
\delta_{\kappa\xi}V^\mu(y) = \kappa \xi^\mu(y)
\eeq
which directly follows  from \eqref{linedress} and the corresponding leading-order transformation\footnote{Interestingly, the dressing \eqref{linedress} can be generalized to involve an integral along an arbitrary line to infinity\cite{QGQF}, and still satisfies \eqref{Kreln} given the following transformation law.}
\beq
\delta_{\kappa\xi}h_{\mu\nu} = -\partial_\mu\xi_\nu -\partial_\nu\xi_\mu\ .
\eeq

An alternative way to write the diffeomorphism-invariant observable $\Phi(y)$ that nicely extends to dressing other observables is\cite{DoGi1,DoGi4}, again to leading order,
\beq\label{Phiexp}
\Phi(y)\simeq e^{i\int d^3x V^\mu(x) T_{0\mu(x)}}\ \phi(y)\ e^{-i\int d^3x V^\mu(x) T_{0\mu(x)}}\ .
\eeq
Here the integral is over a constant time slice, and $T_{0\mu}$ are components of the matter stress tensor, and thus generate translations, reproducing \eqref{Phisum}.

\begin{figure}[h]
 	\begin{center}
 		\includegraphics[width=0.55\textwidth]{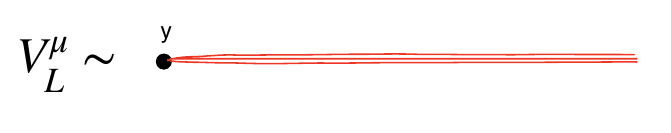} 
 		\caption{Shown is a gravitational line dressing; this may be regulated by averaging over a small cone, or tube.}
 		\label{fig-line}
 	\end{center}
 \end{figure}

In the quantum theory, the operator \eqref{Phisum} or \eqref{Phiexp} acts on the vacuum, or other state, to create a new nontrivial state.  The undressed operator $\phi(x)$ creates or annihilates a particle, and the dressing $V^\mu$ is easily seen to create or annihilate a leading-order gravitational field of that particle\cite{DoGi1}.  This field is in a configuration localized to a line -- or if regulated, a small tube -- extending to infinity, as pictured in Fig.~\ref{fig-line}.

\begin{figure}[h]
 	\begin{center}
 		\includegraphics[width=0.85\textwidth]{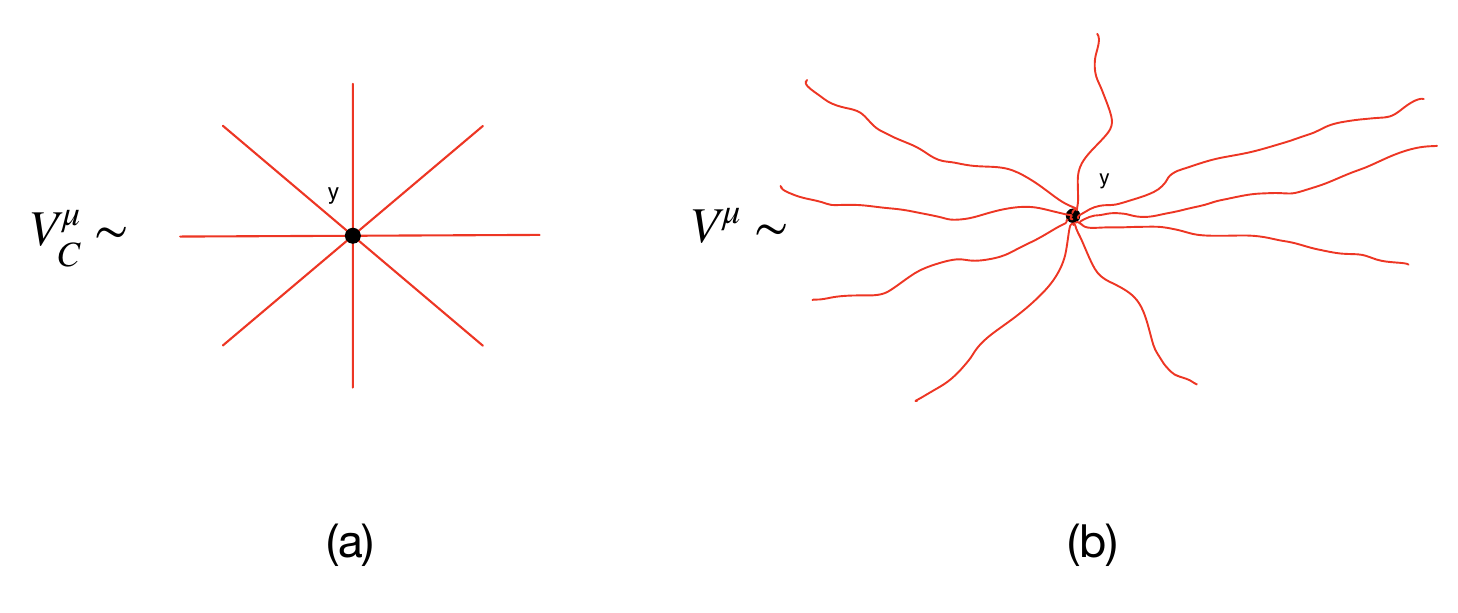} 
 		\caption{Other dressings; (a) shows a Coulomb-like dressing, obtained by uniformly averaging the line dressing of Fig.~\ref{fig-line}, and (b) shows a more general dressing resulting from a non-uniform averaging of more general lines.}
 		\label{fig-coul-gen}
 	\end{center}
 \end{figure}

This construction generalizes in several ways.  First, one may consider more general gravitational dressings -- these are highly non-unique. One approach to this is to average the line dressing $V_L^\mu$ over some range of angles, {\it e.g.} to create a dressing in a cone, or that is spherically symmetric.  The latter dressing $V_C^\mu$ creates a Coulomb-like field, corresponding to linearized Schwarzschild\cite{DoGi1}, as sketched in Fig.~\ref{fig-coul-gen}(a).  Or, one could consider a more complicated integral, creating a non-uniform field, as sketched in Fig.~\ref{fig-coul-gen}(b).  These must all satisfy the key condition \eqref{keyc}.  Moreover, there is a dressing theorem\cite{DoGi2}, stating that such dressings must extend to infinity in flat space, with generalization to other non-compact spacetimes such as AdS\cite{GiKi}.  

Although these constructions are so far only given to leading order in $\kappa$, we expect that they capture the correct long-distance behavior of gravity, since the long-distance behavior generically corresponds to weak-field behavior, suppressing higher-order corrections.  There are questions of higher-order structure, which particularly become important near the origin of the dressing.

Another generalization is that of dressing a more general matter operator $O$ localized to a time slice, where the expression \eqref{Phiexp} nicely extends to\footnote{In principle one could also consider dressing other operators localized to a region, for example those constructed from the metric perturbation $h$ itself, with inclusion of the corresponding stress tensors\cite{GiPe2}.}
\beq\label{dressO}
\hat O\simeq e^{i\int d^3x V^\mu(x) T_{0\mu(x)}}\ O\ e^{-i\int d^3x V^\mu(x) T_{0\mu(x)}}\ .
\eeq

One can also generalize to consider dressings of operators perturbing about more general background metrics\cite{GiPe2}, now with expansion ${\tilde g}_{\mu\nu}=g_{\mu\nu}+\kappa h_{\mu\nu}$ describing the quantum metric.  These can be analyzed in a canonical approach, by choosing a slicing of the spacetime and working in ADM variables\cite{ADM}, with metric
\beq
ds^2=-N^2dt^2+q_{ij}(dx^i+N^i dt)(dx^j+N^j dt)\ .
\eeq
Then the generalization of \eqref{dressO} is
\beq\label{Ogen}
\hat O \simeq e^{i\int d^{3}x \sqrt{q} V^\mu(x)\,T_{n\mu}(x)}\ O\ e^{-i\int d^{3}x \sqrt{q} V^\mu(x)\,T_{n\mu}(x) }\ ,
\eeq
with a general dressing
\beq
V^\mu(x) = \kappa \int d^{3}x'\left[ G^{\mu ij}(x,x') \ h_{ij}(x') + H^\mu_{ij}(x,x')\ p^{ij}(x') \right]\ .
\eeq
In this expression $p^{ij}$ is the canonical conjugate to $h_{ij}$, satisfying the equal time commutators
\beq
[p^{ij}(x),h_{kl}(x')]= -\frac{i}{2}\left(\delta^i_k \delta^j_l+\delta^i_l \delta^j_k\right) \delta^3(x-x'),
\eeq
and the functions $G^{\mu ij}$ and $H^\mu_{ij}$ are solutions to a generalized Green function problem for linearized perturbations about the background metric, described in \cite{GiPe2}.\footnote{The basic equations appear as (4.17)-(4.20) of that reference, but due to their somewhat complicated detail are not reproduced here.}  Notice that in \eqref{Ogen}, the exponentials take the form of Weyl operators for the Hilbert space of metric perturbations.

In fact, an alternate way to derive \eqref{Ogen}  and the corresponding Green function equations is to use the fact that the constraints
\beq
{\cal C}_\mu(x)=\left[-\frac{G_{\mu\nu}(x)}{8\pi G}+T_{\mu\nu}(x)\right]\sqrt q n^\nu\ ,
\eeq
 with $G_{\mu\nu}$ the Einstein tensor and $n^\mu$ the normal to a time slice, generate diffeomorphisms (on shell), and so a diffeomorphism-invariant operator should commute with them,
\beq
[{\cal C}_\mu(x),\hat O]=0\ .
\eeq
Solving this condition to leading order in $\kappa$ gives the Green function equations and \eqref{Ogen}.

Here, again, we encounter a high degree of nonuniqueness, now associated with the infinite number of possible solutions of the Green function equations, depending on the boundary conditions.  The difference between such solutions is of course associated with a solution to the homogeneous problem.  As a result, the difference, {\it e.g.} of the general dressing $V^\mu$ of Fig.~\ref{fig-coul-gen}(b) and the Coulomb dressing $V_C^\mu$  of Fig.~\ref{fig-coul-gen}(a), corresponds to a homogeneous solution of the linearized field equations, that is to a radiative or propagating gravitational field.  Such a field may be added to a given dressing to give a different dressing, and the general such field decays by radiating to infinity, leaving a Coulomb-like configuration\cite{DoGi1}.

\section{Information localization and gravitational splitting}
\label{secgrsp}

We reiterate that such constructions of gravitationally-dressed observables are minimal, in the sense that they involve no extra structure beyond the gravitational dressing required for gauge invariance, corresponding to the gravitational field of the matter.  But, given the key role of observables for information localization  in other quantum theories such as LQFT, this then
raises a very important question -- how is information now localized in quantum gravity?

\begin{figure}[h]
 	\begin{center}
 		\includegraphics[width=0.70\textwidth]{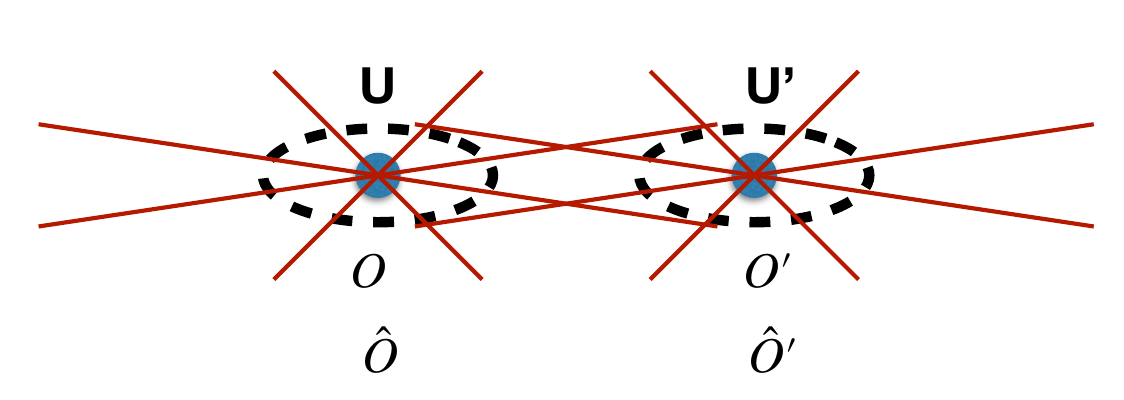} 
 		\caption{Illustration of the general noncommutativity of gravitationally dressed observables associated with spacelike separated regions $U$ and $U'$; the dressing parts of the operators lead to nontrivial commutation relations.}
 		\label{fig-noncomm}
 	\end{center}
 \end{figure}

\begin{figure}[h]
 	\begin{center}
 		\includegraphics[width=0.60\textwidth]{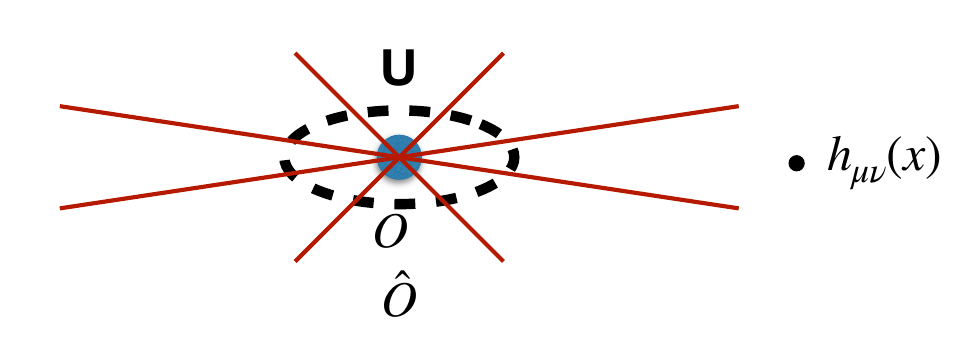} 
 		\caption{Schematic of measurement of a dressed operator based in a region $U$ through measuring the gravitation perturbation at a spacelike separated point.}
 		\label{fig-hmeas}
 	\end{center}
 \end{figure}

We reviewed the role of the algebra of observables of LQFT in Sec.~\ref{LQFTsec}.  With the gravitationally-dressed observables in hand, one can readily see that the dressing significantly changes the algebra\cite{SGalg,DoGi1}.  This is illustrated in Fig.~\ref{fig-noncomm}, for dressed operators such as \eqref{dressO}.  The dressing part of the operator extends to infinity, and so two such operators based on generic spacelike separated underlying LQFT operators $O$, $O'$ will generically not commute, $[\hat O,\hat O']\neq 0$, due to the noncommutativity of this dressing\cite{SGalg,DoGi1}.  Since operators spacelike to $O$ no longer commute with $\hat O$, this suggests that observations spacelike separated from $U$ can gain information about the state created by $\hat O$; as a simple example, we might think that we could gain such information by measuring the metric perturbation $h_{\mu\nu}$, as illustrated in Fig.~\ref{fig-hmeas}.  Specifically, if we consider a state created by a localized source $J$,
\beq
{\hat {| \psi_J\rangle}}=e^{-i\int J\hat O}|0\rangle\ ,
\eeq
we for example expect that
\beq
\langle\psi_J|h_{\mu\nu}(x)|\psi_J\rangle\neq \langle0|h_{\mu\nu}(x)|0\rangle\ ,
\eeq
in contrast to the result  \eqref{statloc} in LQFT.

However, this raises the question of how much information about $O$ is actually accessible by such separated measurements.  And, at least at the perturbative level, there are limitations to this that arise from considering what has been called\cite{DoGi3,DoGi4,SGsplit} the ``Standard dressing" construction, and the resulting notion of a perturbative ``gravitational splitting."

\begin{figure}[h]
 	\begin{center}
 		\includegraphics[width=0.45\textwidth]{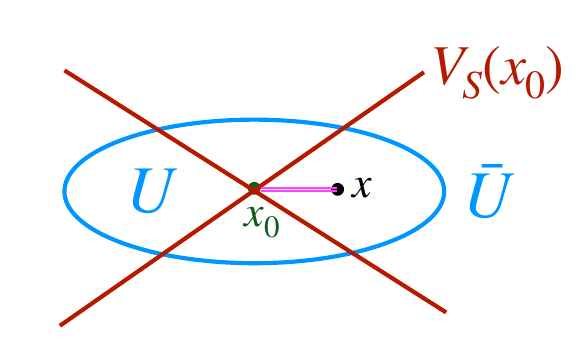} 
 		\caption{Illustration of the standard dressing construction.  An operator at a general $x\in U$ can be dressed to a specified point $x_0$, which in turn has a standard dressing extending to infinity.}
 		\label{fig-sdress}
 	\end{center}
 \end{figure}

The basic idea behind these goes back to the nonuniqueness of the dressing $V$, described in the preceding section.  This means that the operator $O$ does not uniquely determine an observable gravitational field outside the the region $U$ in which it is localized; much of the information in that field is determined, at least at the perturbative level, by gravitational degrees of freedom outside $U$.  To explore how much information is ``expressed" by the field outside $U$, we can begin with $U$ and a chosen point $x_0\in U$, and construct a {\it standard dressing} $V_S(x_0)$ associated with $U$ and that point, and specifically satisfying the key relation \eqref{Kreln} at  $x_0$,
\beq
\delta_{\kappa\xi}V^\mu(x_0) = \kappa \xi^\mu(x_0)\ .
\eeq
For example, $V_S(x_0)$ could be either a line dressing or a Coulomb dressing, or one of another form, anchored at $x_0$.  Then, the dressing of a general point $x$ can be constructed by constructing a dressing $V^\mu(x,x_0)$ between $x$ and $x_0$; at the leading level an example is the line dressing \eqref{linedress}, now with integration between upper and lower limits $x_0$ and $x$, respectively, or one could consider generalizations.  Combining these results in a dressing for a general point $x$,
\beq
\label{Sdress}
V^\mu(x) = V^\mu(x,x_0) + V^\mu_S(x_0) +\frac{1}{2}(x-x_0)_\nu\left[\partial^\nu V_S^\mu(x_0) - \partial^\mu V_S^\nu(x_0)\right]\ ,
\eeq
where the last term is easily seen to be needed so that $V^\mu(x)$ satisfies the key relation \eqref{Kreln}.  This construction is illustrated in Fig.~\ref{fig-sdress}.

Given this construction, now consider the localized LQFT state $|\psi_J\rangle=U_J|0\rangle$ of \eqref{Jstate}, with
\beq
U_J= e^{-i\int dx J(x) O(x)}\ ,
\eeq
and its dressed version
\beq
\hat{|\psi_J\rangle} = e^{i\int d^3x V^\mu(x) T_{0\mu(x)}} |\psi_J\rangle
\eeq
with $V^\mu$ of \eqref{Sdress}.  Then, working to leading order in $\kappa$, we find that for $x$ outside $U$,
\bea
\label{stanmeas}
\hat{\langle \psi_J|}h_{\mu\nu}(x){\hat {| \psi_J\rangle}}&=&- i \langle 0| U_J^\dagger\, [P_\lambda, U_J]|0\rangle\ \langle 0| [h_{\mu\nu}(x),V^\lambda_S(x_0)]|0\rangle \cr &&-\frac{i}{2} \langle 0| U_J^\dagger [M_{\lambda\sigma} ,U_J]|0\rangle\ \langle 0| [h_{\mu\nu}(x),\partial^\lambda V_S^\sigma(x_0)]|0\rangle
\eea
where $P_\mu$ and $M_{\mu\nu}$ are the Poincar\'e generators,
\beq
P_\mu=-\int d^3x T_{0\mu}(x)\quad ,\quad  M_{\mu\nu}=-\int d^3x\left[(x-x_0)_\mu T_{0\nu}(x) - (x-x_0)_\nu T_{0\mu}(x)\right]\ .
\eeq

In each of the two terms of \eqref{stanmeas}, the first factors describe Poincar\'e moments of the state created by $U_J$.  The second factors describe the gravitational field created by the standard dressing $V_S(x_0)$ of the chosen point.  Correspondingly, the measurement of $h_{\mu\nu}$ outside $U$ only depends on the state $|\psi_J\rangle=U_J|0\rangle$ through its Poincar\'e moments, and different states with the same Poincar\'e moments are not distinguished by such measurements, at this leading order in $\kappa$. 

This construction was called a gravitational splitting in \cite{DoGi4}.  It gives a quantum extension of classical results of \cite{CaSc,Chru,CoSc}, which show that the classical gravitational field of localized matter may be taken to be of the form of the boosted Kerr solution, or of a conically-localized form, generalizing the localization of the gravitational line \eqref{linedress}.  
To understand the result \eqref{stanmeas}, note first that with \eqref{Sdress}
\beq
\int d^3x V^\mu(x)\, T_{0\mu(x)} = - P_\mu V_S^\mu(x_0) -\frac{1}{2}M_{\mu\nu}\partial^\mu V^\nu_S(x_0) + \int d^3x V^\mu(x,x_0) T_{0\mu}(x)\ ,
\eeq
and then that for $O$ supported only in $U$, and to leading order,
\beq\label{dressp}
{\hat O}\simeq e^{i\int d^3x V^\mu(x) T_{0\mu(x)}}\,O\,e^{-i\int d^3x V^\mu(x) T_{0\mu(x)}}  \simeq \ e^{- iP_\mu V_S^\mu -\frac{i}{2}M_{\mu\nu}\partial^\mu V^\nu_S}\,{\tilde O}\, e^{ iP_\mu V_S^\mu +\frac{i}{2}M_{\mu\nu}\partial^\mu V^\nu_S}\ ,
\eeq
with
\beq\label{dresto0}
{\tilde O} \simeq e^{i\int d^3x V^\mu(x,x_0) T_{0\mu}(x)}\,O\,e^{-i\int d^3x V^\mu(x,x_0) T_{0\mu}(x)}\ ;
\eeq
we can think of the latter as the observable $O$ dressed to the chosen point $x_0$.
This $\tilde O$ is also localized to $U$ if we use a $V^\mu(x,x_0)$ that stays localized to $U$ for $x\in U$.  

\section{Algebraic structure and connection to the crossed product and type II algebras}
\label{crosssec}

The gravitationally-dressed observable \eqref{dressp} gives a {\it generalization} of the crossed product construction\cite{Take}, which has been of considerable interest recently in beginning to understand gravitational modifications of the algebraic structures of LQFT\cite{Wittcross}\cite{CLPW,CPW}.  
If $\cala$ is an algebra and $G$ is a group of outer automorphisms of $\cala$, the crossed product  ${\cal A}\rtimes G$ is generated by elements
\beq\label{crossdef}
\{e^{-ix_aH_a}\, {\bf a}\, e^{ix_aH_a}, p_a\}
\eeq
where $\bf a$ is an element of $\cala$, $x_a$ may be thought of as a coordinate on $G$, $H_a$ is a generator of the action of $G$ on $\cala$, and $p_a$ are conjugate variables to the coordinates $x_a$.\footnote{Here we differ in notation with \cite{CLPW,CPW,WittARO}, who denote what we call $x$ by $p$.}
The expression \eqref{dressp} at first  looks like it describes such a crossed product of the subalgebra of observables $\cala_U$ associated with region $U$ with the Poincar\'e group.  For example, one readily finds\cite{DoGi1} that the ADM momenta $P_\mu^{ADM}$ satisfy
\beq
[P_\mu^{ADM},V_S^\nu] = i\delta_\mu^\nu\ ,
\eeq
so $P_\mu^{ADM}$ and the standard dressing $V_S^\nu$ are conjugate variables, directly analogous to the $p_a$ and $x_a$, respectively, in \eqref{crossdef}, with a similar apparent structure involving the generators $M_{\mu\nu}$ and the corresponding ADM expressions.
However, \eqref{dressp} represents a more general construction since the generators of the Poincar\'e group are not in general automorphisms of $\cala_U$ -- these generators typically ``move the region $U$," and so do not map $\cala_U$ to itself.

\begin{figure}[h]
 	\begin{center}
 		\includegraphics[width=0.70\textwidth]{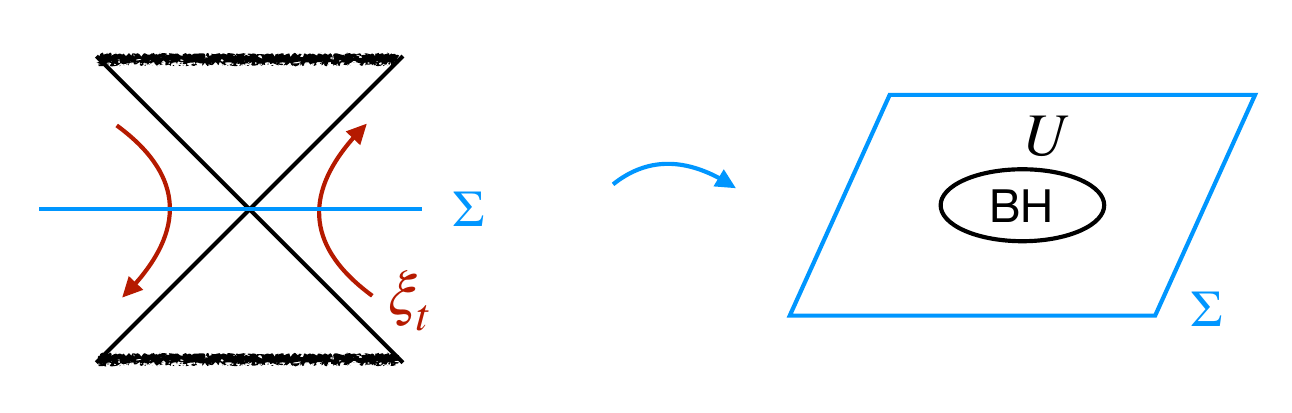} 
 		\caption{The spacetime of an eternal black hole, and a spatial slice, for which the region interior to the horizon is treated as our region $U$.}
 		\label{fig-bh}
 	\end{center}
 \end{figure}

This discussion suggests a way to find special cases where one can reproduce the crossed product, and then {\it e.g.} make direct connection to the construction of type II von Neumann algebras\cite{CPW,KLS}.  
Suppose for example one works about a black hole spacetime, as illustrated in Fig.~\ref{fig-bh}, and that there is a standard dressing construction generalizing  \eqref{dressp} to this background.\footnote{While we do not give such an explicit dressing here, we note the relation of its construction to the problem of finding linearized perturbations of the gravitational field\cite{GiPe2}, as described in Sec.~\ref{GDO}, and expect that we can find such perturbations carrying the asymptotic Poincar\'e charges.  Some ingredients for this appear also in \cite{GiWe}.}  Specifically, consider the time translation generator $P_t$, corresponding now to the Killing vector $\xi$ of the BH geometry, and its associated standard dressing $V^t_S$, which asymptotically should behave like that of a flat space region.  In this case, $P_t$ is an automorphism of the algebra of observables associated with the region $U$ which are ``interior to the BH" as viewed from the right asymptotic observer.  Specifically, the generator is
\beq
-P_t=\int_\Sigma d\Sigma\, n^\mu \xi_t^\nu T_{\mu\nu} = H
\eeq
which is the modular hamiltonian for the Hartle-Hawking state of the BH, and correspondingly $P_t$ generates a ``modular automorphism" of $U$.  
If we consider a {\it truncated} dressed observable
\beq\label{truncobs}
\hat O_{trunc} = e^{-i V^t_S P_t}\, O \,e^{i V^t_S P_t}\ 
\eeq
(or with $O$ dressed to an internal point, as in \eqref{dresto0}; compare \eqref{dressp})
this commutes with $H^{ADM}+P_t$, {\it i.e.} with the leading-order constraint $C_\xi$ where $\xi$ is the Killing vector.  It does not, however, in general commute with $\calc_\mu(x)$, due to the truncation; the full dressing is needed to commute with the full set of constraints, as studied for example in \cite{GiPe2}.  A construction like \eqref{truncobs}  is what was used in \cite{CPW} to show that the type III von Neumann algebra $\cala_U$ becomes a type $II_\infty$ algebra, allowing the definition of an entropy.\footnote{One could also consider dressing of operators involving the metric perturbation, as noted earlier.}

\begin{figure}[h]
 	\begin{center}
 		\includegraphics[width=0.45\textwidth]{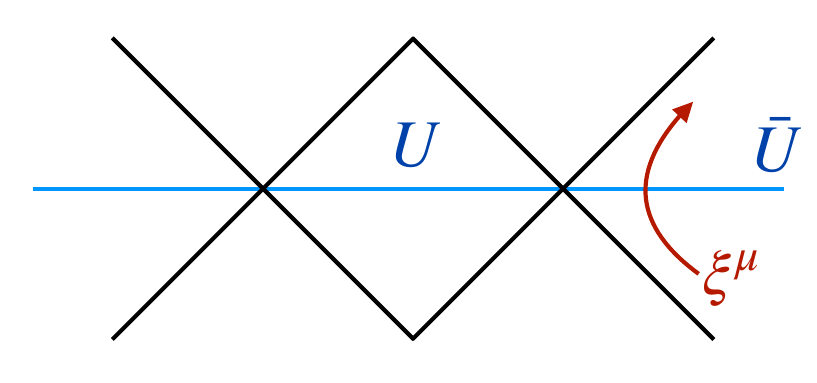} 
 		\caption{Illustration of a flat spacetime diamond, and corresponding Killing vector leaving it fixed.}
 		\label{fig-flat}
 	\end{center}
 \end{figure} 

\begin{figure}[h]
 	\begin{center}
 		\includegraphics[width=0.35\textwidth]{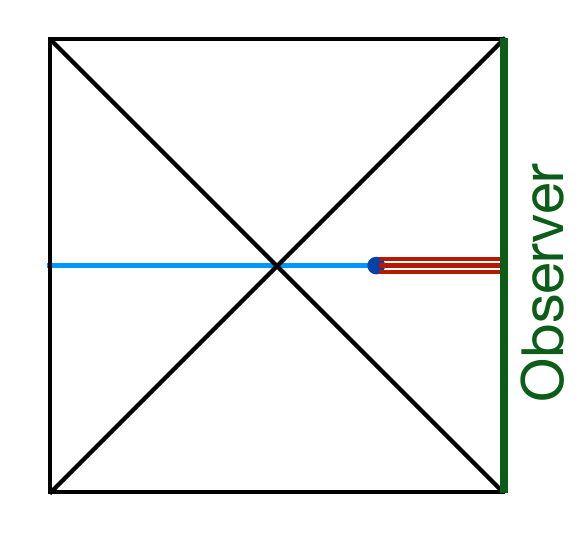} 
 		\caption{Illustration of an observable in de Sitter space, constructed by gravitationally dressing to a ``central observer."}
 		\label{fig-ds}
 	\end{center}
 \end{figure}

There is a similar story for subregions $U$ of flat space\cite{JSS}, as illustrated in Fig.~\ref{fig-flat}, with a similar potential issue with solving all the constraints.  Also, if one considers observers in de Sitter space, there is a related story describing recovery of type II algebras, which one can think of as involving field relational, or more generally mixed, observables which include some form of observer, as is  illustrated in Fig.~\ref{fig-ds}, and discussed beginning in \cite{CLPW}; see \cite{ChPe,KLScosmo,DEHK-L} for further discussion.  Similar constructions appear relevant for observers {\it within} BH spacetimes.

\section{The question of algebraic or more general mathematical structure}

The gravitational transition of type III algebras to type II, arising {\it e.g.} from the truncated dressed observables \eqref{truncobs}, has been argued to be an important one.  However, we have just seen it is just a small part of a {\it more profound modification} of the algebraic structure, associated with quantum gravity.  One apparently important aspect of this modification is that we no longer have the same kind of commuting subalgebras; whereas the quantum field theory subalgebras ${\cal A}_U\ ,\ {\cal A}_{U'}$ associated to spacelike-separated regions commute, the algebras of the corresponding dressed observables ${\widehat {\cal A}}_U\ ,\ {\widehat {\cal A}}_{U'}$ generically do not\cite{SGalg,DoGi1}, as illustrated in Fig.~\ref{fig-noncomm}.

This means that we have a very different algebraic structure associated with regions.  Since the algebras of these regions no longer commute, that leaves us with a key question:  how do we define subsystems in the quantum-gravitational context?  As we have described, this connects directly to the question of how we think about Einstein separability -- which was argued to be an essential part of describing physics.

In fact, the gravitational splitting construction of Sec.~\ref{secgrsp} is not expected to precisely localize information.  Suppose, going beyond the construction of the leading dressing, we are able to solve 
\beq
[{\cal C}_\mu(x),\hat O]=0
\eeq
to {\it all} orders, to construct fully gravitationally dressed observables, {\it e.g} in a nonlinear generalization of the standard dressing construction.  Then, if $\hat O$ is based in some region, we expect that
\beq
e^{-ic^i P_i^{ADM}}\, \hat O\, e^{ic^i P_i^{ADM}} = \hat O_{c^i}
\eeq 
is based in the translate of that region by $c^i$, since 
\beq
P_i^{ADM}= P_i + \int d^3x\, {\cal C}_i
\eeq
with $P_i$ the translation generators, and since $\hat O$ commutes with the $\calc_i$.  This would imply that we can act with {\it asymptotic operators} -- the ADM momenta, which are defined as integrals of the metric perturbation $h_{ij}$ at infinity -- to shift the operator $\hat O$ over a long distance, {\it e.g.} so that it approaches a boundary of spacetime\cite{DoGi3}.  For example, in an analogous construction in AdS, generalizing \cite{GiKi}, one might shift the operator to approach the AdS boundary.  Indeed, this seems to lead to an argument for ``gravitational holography" that is similar in spirit to Marolf's argument\cite{Maroholo}, also using the constraints, but even simpler.  Of course, important questions remain, such as the higher-order construction of the dressing, and the nonperturbative completion of this story.\footnote{Even a leading perturbative version of these effects suggests   sensitivity of distant operators to excitations in a region\cite{Rajuetal,SGsub}; however, such effects for massive excitations appear exponentially small in  distance\cite{SGsub}.}

\begin{figure}[h]
 	\begin{center}
 		\includegraphics[width=0.60\textwidth]{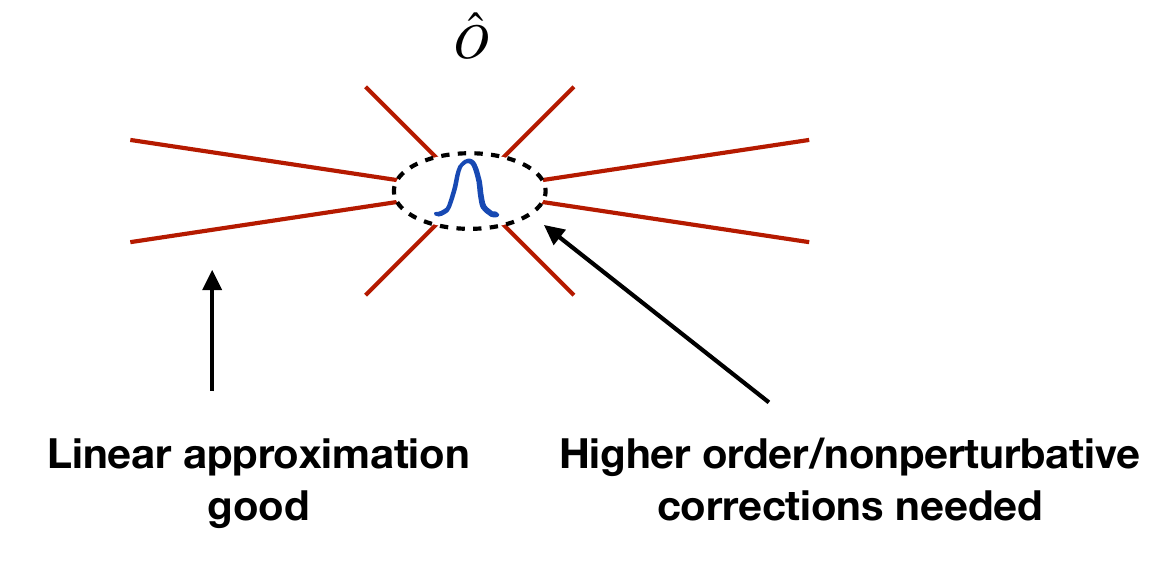} 
 		\caption{Illustration of the dressing of an operator creating a localized wavepacket.  At large distances this is well-approximated by the leading dressing, but at short distances important nonlinear corrections are expected.}
 		\label{fig-nonlin}
 	\end{center}
 \end{figure}

Indeed, the latter illustrates even deeper apparent subtleties in localizing information, since the expected nonperturbative structure suggests even more radical behavior, which might be called ``algebraic spacetime disruption."  Consider a fully-dressed observable $\hat O$ associated with an underlying operator that is supposed to create a localized wavepacket of matter, as illustrated in Fig.~\ref{fig-nonlin}.  Since the gravitational field weakens asymptotically, we expect the leading dressing constructions described earlier to be valid asymptotically, but at shorter distances higher order and/or nonperturbative corrections are needed.  Then, if the relevant structure is that of an algebra $\hat \cala$ of such operators, we should be able to define operators $\hat O^N$.  As $N$ grows, the region of nonlinear gravitational dressing grows; we expect that the operator does something like create a big quantum black hole.  And, in the limit $N\rightarrow\infty$, this phenomenon consumes the entire spacetime.

These problems of localizing information and spacetime disruption seem to suggest that we need a different, {\it non-algebraic mathematical structure} on the Hilbert spaces describing quantum gravity.  A big question is:  what is it? One initial idea, very preliminarily discussed in  \cite{QFG,QGQF,SGsplit}, is that a more limited structure than operator subalgebras is a network of Hilbert space inclusions, that serves to localize information.  Further exploration of the relevant mathematical structures is clearly a very interesting problem for the future.

In conclusion, the questions surrounding observables and observation are key questions in quantum gravity, with different facets.  These include what observers actually measure, what can in principle be measured, and certain operators of the quantum theory thought of as observables.  Perturbatively one can construct such operators by working relationally.  Extending what we know about local QFT, we expect such observables to be closely connected to the fundamental structure of the theory.  However, we seem to find that this structure is profoundly different from that of LQFT; the story of how type III algebras become type II is just a small piece of this.  And, when one considers the expected behavior of the full algebras, that suggests that other mathematical structure on Hilbert spaces for quantum gravity may be an essential part of its definition.

\vskip.3in
\noindent{\bf Acknowledgements} 

This material is based upon work supported in part by the U.S. Department of Energy, Office of Science, under Award Number {DE-SC}0011702, and by Heising-Simons Foundation grants \#2021-2819 and \#2024-5307.  I thank D. Marolf for helpful comments on a preliminary draft.

\mciteSetMidEndSepPunct{}{\ifmciteBstWouldAddEndPunct.\else\fi}{\relax}
\bibliographystyle{utphys}
\bibliography{qgobs}{}

\end{document}